\documentclass{evn2004}
\usepackage{txfonts}
\usepackage{graphicx}
\begin{document}
   \title{Radio-loud and Radio-quiet X-ray Binaries: LSI+61$^{\circ}$303 in Context}
   \author{Maria Massi}
   \institute{
Max Planck Institut f\"ur Radioastronomie, Auf dem H\"ugel 69,
D-53121 Bonn, Germany}
   \abstract{
The three basic ingredients - a spinning compact object, an accretion
disc and a collimated relativistic jet - make microquasars  a galactic
scaled-down version of the  radio-loud AGN.  That explains the
large interest attributed  to this new class of objects, which up to now
consists of less than 20 members. Microquasars belong to the much
larger class of  X-ray binary systems,
where there exists a compact object  together with its X-ray emitting
accretion disc, but the  relativistic jet is missing.
When does an X-ray binary system evolve into a microquasar?
Ideal for studying such  kind of a transition is the periodic microquasar
LS~I~+61$^{\circ}$303  formed by  
a  compact object  accreting  from the equatorial wind of
a Be star and with  more than one  
 event of  super-critical accretion and ejection  along the eccentric
orbit.
For ejections at periastron passage the relativistic electrons 
suffer  severe  inverse Compton losses by upscattering
the UV photons of the Be star  at high energy 
: At periastron passage Gamma-ray emission has been observed,
whereas radio outbursts have never been observed in 20 years of radio flux monitoring.
For ejections displaced from periastron passage 
the losses are less severe and radio outbursts are observed.
The radio emission   mapped   
on scales  from a few AU  to hundreds of AU 
shows a double-sided relativistic ($\beta=0.6c$) S-shaped jet, 
similar to the  well-known precessing jet of \object{SS~433}.
   }

   \maketitle
\section{Introduction}
Since the beginning of the 1980s radio-galaxies,  quasars, Seyferts, QSO  etc.
all are simply classified as  AGNs (``Active Galactic Nuclei'')
because the ``energy-engine'' is thought to be the same:
A super-massive black hole accreting from its host galaxy.
AGNs having  radio-emitting lobes or jets are
called radio-loud, the others
are called radio-quiet 
(Ulrich et al. 1997).

In  an X-ray binary system the ``energy-engine'' 
 is  a compact object of a few
solar masses accreting from the companion star.
Up to now there are almost 250  known X-ray binaries (Liu 2000).
Only a small percentuage of them ($<$10\%) show evidence of a radio-jet
and therefore are radio loud applying the same definition as for the AGNs.
The radio loud X-ray binaries  subclass (Fig. 1) 
includes together with the microquasars
--objects where high resolution radio interferometric techniques
have shown the presence of collimated  jets (Mirabel et al. 1992)--
also   unresolved radio sources with a flat spectrum. 
This  spectrum
 can arise from the combination of emission from optically thick and thin regions
 of an  expanding  continuous jet
(Hjellming \& Johnston 1988;   Fender 2004) as  has been shown 
by the discovery of a continuous   jet for the flat-spectrum source Cygnus X-1
(Stirling et al. 2001).
   \begin{figure*}
   \centering
\includegraphics[width=15cm]{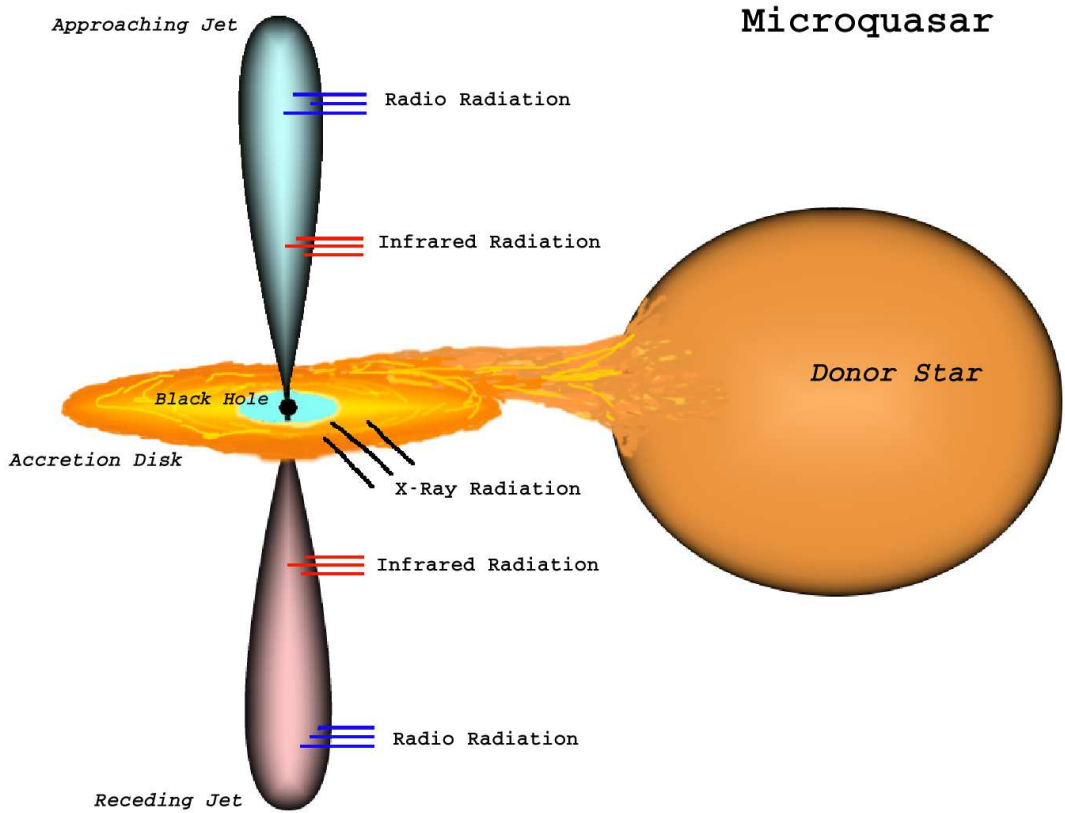}
   \caption{
The   basic components of a microquasar
 - a spinning compact object, an accretion disk and a collimated
 relativistic jet.
 The compact object, of a few solar masses,   accretes from a normal star
in orbital motion around 
  it.  The mass
    of the compact object can be  determined by studying  the   
periodical
  shift of the optical spectral lines
  of the normal star  and   establishes, if it is a neutron star or a black hole.
    The inner part of the disk
     emits X-rays. The inner radius, three times the Schwarzschild radius,
is a few tens of kilometers,
     the outer radius a factor 10$^3$ larger (the figure is not to scale).
     Due to magneto-rotational instabilities part of the disk is propelled
     into a relativistic jet, studied at high resolution with radio
     interferometric techiques.
     In some microquasars, like SS433 and LS I+61$^{\circ}$303, the jet
      is precessing. If the precession causes the jet to be aligned
 toward the earth
      the large variable doppler boosting  mimics the variability of  Blazars
      and the microquasar in this case is called microblazar.
           }
    \end{figure*}

Besides the first radio loud  X-ray binary system  
SS433, discovered by
chance in 1979 (Margon 1984),
 the others have been discovered mostly in the last ten years
and now  are considered as an ideal, nearby  laboratory for
studying the processes of accretion and ejection around black holes.

More generally speaking, the class of the 
X-ray binaries is formed by  stellar systems of two stars
with very different natures:  a normal star (acting as mass donor)
 and a compact object (the accretor) that can
be  either a neutron star or a black hole (White et al 1996).
The normal star  orbits around the compact object and therefore
the infalling material
has some angular momentum ($J$), which prevents  it  from
falling directly into the accretor.
The stream of matter orbits the compact object with a radius
determined  by  $J$  and  the  mass of the compact object ($M_x$).
The angular momentum is  redistributed by the viscosity:
Some of the material takes angular momentum  and
spreads outwards, whereas other material spirals inwards. In
this  way a disk is created from the initial ring of matter
(Longair 1994; King 1996).
Gradually,  the matter drifts inwards until it reaches the last stable orbit,
The viscosity has two effects: Beside the transport of the angular-momentum 
it also acts as a frictional force resulting in the dissipation of heat.
The amount of friction depends
on how  fast the gas orbits around the compact object.
It reaches its maximum at the inner disk, where
the matter is heated up to quite high temperatures (tens of millions of  degrees),
 producing the strong  thermal X-ray radiation
giving the  name ``X-ray binaries'' to this class of objects.

In the case of a low vertical magnetic field threading the disk
the  plasma pressure dominates the magnetic field pressure and
the differentially rotating  disk bends
the magnetic field, which is passively wound up
(Meier et al. 2001).
Due to the compression of the magnetic field lines
the magnetic pressure may  become larger than the gas pressure at
the surface of the accretion disk, where the density is lower.
At this point the gas follows the twisted magnetic field lines,
creating  two spinning flows.
These extract angular momentum 
 from the surface of the disk (magnetic braking) and enhance the radial accretion.
 The avalanching material further pulls
the deformed magnetic field with it
and afterwards magnetic reconnection may happen (Matsumoto et al. 1996).
The thickness of the disc is fundamental in this  magneto-rotational process,
or better the extent of the poloidal magnetic field frozen into the disc   
(Meier 2001; Meier et al. 2001; Maccarone 2004).
No radio jet is associated at  X-ray binaries in  
High/soft states, where the X-ray spectrum is dominated by a
 geometrically thin (optically thick)
 accretion disc (Shakura \& Sunyaev 1973).
Whereas, numerical results show a jet being launched 
from an  inner geometrically
thick portion of the accretion disc existing (coronal flow/ADAF)
when the   X-ray binaries are in their low/hard state
(Meyer et al. 2000; Meier 2001).

Among the X-ray binaries an ideal source to study the transition to the
microquasar phase is  LS~I~+61$^{\circ}$303 because it is the only 
known periodic microquasar.

\section {The LS~I~+61$^{\circ}$303 system}

LS~I~+61$^{\circ}$303 is the only object of
this class showing variations in  X-rays
(Leahy 2001),
optical wavelenghts in both continuum (Maraschi \& Treves  1981)
and line radiation (Zamanov \& Mart\'{\i}  2000; Liu et al. 2000)
and at radio wavelenghts
with a period equal to the orbital one;
the most accurate value of the orbital period is that resulting
 from radio observations, equal to
26.496 days (Gregory \& Taylor 1978;
Taylor \& Gregory  1982; Gregory  2002).
The fit performed on near infrared data  by Mart\'{\i} and Paredes
(1995)
 produced  high  values  for the  eccentricity (e$\sim$ 0.7-0.8 )
confirmed by optical observations
(Casares et al. 2004).
The  lower limit to  $i$,  
the angle formed by the axis of the orbit and the line of sight 
  is  38$^{\circ}$
 (Hutchings \& Crampton 1981; Massi et al. 2001; Massi 2004).
The  phase at the periastron passage
is $\Phi$=0.2 
with the phase  referred to the time t$_0$=JD\,2443366.775,
the date of the first radio detection of the system
(Gregory \& Taylor 1978).

Ultraviolet spectroscopy of LS~I~+61$^{\circ}$303  by Hutchings \&
Crampton (Hutchings \& Crampton  1981)
 indicates that the normal star is a main sequence B0-B0.5 star (L$\sim 10^{38}$
 erg sec$^{-1}$, T$_{eff}\simeq 2.6~10^4$K).
The optical spectrum is that of a rapidly 
spinning Be star.
Together with the usual  high velocity  (1000 km s$^{-1}$)
 low density wind at high latitudes typical for OB stars,
Be stars have a dense and slow ($<$100 km~ s$^{-1}$) disk-like
wind around the equator (Waters et al. 1988).
Equatorial mass loss, due to an interplay of the high rotation and
of internal pulsations of the star,  is highly variable
and in  some cases  periodical.
% with periods of years.
This is the case for LS~I~+61$^{\circ}$303, where 
periodical variations 
of the mass loss from the Be star
have been proved  
by a  modulation of  H$\alpha$ emission line
with a period of almost 4 years
(Zamanov and Mart\'{\i} 2000).

The most reliable method to determine the nature of
the compact object is the study, as usual  in binary  systems,
 of the changing radial velocity of the normal companion during its orbit.
The   amplitude  ($K_c$) of the  radial velocity variations 
 and the period (P$_{orb}$) of the system
define a quantity, called the ``mass function'' (Charles \& Wagner 1996),
$f$, which depends on the inclination $i$  of the orbit, the masses $M_X$ and $M$
of the accretor and its normal companion:
$$f ={P_{orb} K_c^3\over 2\pi G}={M_X^3 sin^3 i\over (M_X +M)^2} $$
where  $G$ is the gravitational constant.

Once the inclination, $i$,
 and the mass of the companion, $M$, are known one
can solve for  $M_X$.

 Rhoades \& Ruffini (1974)
by taking the most extreme equation of state that produces 
the maximum critical mass of  a neutron star, established the upper
limit of 3.2 $M\odot$.
This absolute maximum mass  provides a decisive method
of observationally distinguishing neutron stars from black holes.
The problem with LS~I~+61$^{\circ}$303 is the large range for the mass function
allowed by the parameter uncertainties.
Optical observations (Casares et~al.
2004) give the mass function $f$ in the range $0.003<f<0.027$.
The upper limit for $f$, with i=38 and  $M = 18\,M_\odot$
gives  $M_{\rm X}<3.8\,M_\odot$.  
Therefore  the nature of the accretor is still  an open issue
(Massi 2004).

\section {\bf Radio emission}

\begin{figure}[htb]
\centering
\resizebox{\hsize}{!}{\includegraphics[scale=0.15, angle=0]{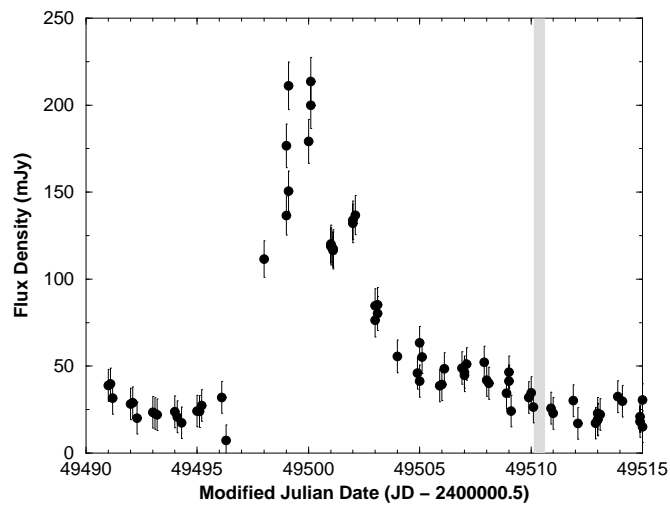}}
\caption{
Radio light curve of LS~I~+61$^{\circ}$303 obtained with the GBI at 8.4~GHz (Strickman et~al. 1998). The shaded area indicates the time interval during the EVN  observation  of Fig. 4
(figure in  Massi et al. 2001).}
\label{fig:flusso}
\end{figure}

The greatest  peculiarity of LS I +61$^{\circ}$303 are
its periodic radio outbursts  with P=26.496 days 
(Gregory 2002).
In Fig.\ref{fig:flusso} a typical radio light curve  is shown.
The decay of the outburst agrees with that expected for an adiabatically 
expanding cloud of synchrotron-emitting relativistic electrons (Taylor \& Gregory 1984).
However,  that  model alone
fails to fit the peaks at  different frequencies: the flat spectrum during the late portion of the rise in flux
density can be reproduced , if together
with the adiabatic expansion losses 
also  a continuos ejection of particles lasting two days is taken into account
(Paredes et al 1991).

The amplitude of each outburst is not randomly varying, but itself periodic
with a periodicity of 4.6 years  
correlated with  the mass loss of the Be star (sect. 2) (Gregory 1999, 2002;
Zamanov \& Mart\'{\i}  2000).
Also, the orbital phase
 $\Phi$ at which the
outbursts occur is modulated (Gregory et al. 1999) and varies   in the interval
0.45--0.95 (Paredes et al. 1990) 
The  $\Phi$ at periastron passage is   0.2. Therefore
one of the fundamental questions concerning the periodic radio
outbursts of \object{LS~I~+61$^{\circ}$303}  has been:
Why are the radio outbursts shifted with respect to the periastron passage?

\section {\bf The two peak accretion model}

In order to explain the association between LS~I~+61$^{\circ}$303 and the
gamma-ray source 2CG 135+01/3EG J0241+6103,
 Bosch-Ramon and Paredes (2004) have 
proposed a numerical  model  based on inverse Compton scattering.  
The  relativistic electrons  in the jet are  exposed to  stellar  photons (external Compton)
as well as to synchrotron photons (synchrotron self Compton).
The model considers  accretion variations along the orbit and predicts
a gamma-ray peak at periastron passage where the accretion 
is higher.
EGRET data show indeed a  gamma-ray peak at  $\Phi$=0.2 (Fig. 3) and 
catastrophic inverse Compton losses might explain the absence of radio emission at periastron.
Therefore, to explain the observed periodic  
 radio outbursts in the phase interval
0.45--0.95  
a second accretion/ejection event must occur. 

Taylor et al. (1992) and Mart\'{\i} \& Paredes (1995) have shown that for 
accretion along an eccentric orbit 
the accretion rate $\dot{M} \propto {\rho_{\rm wind}\over v_{\rm rel}^3}$, (where $\rho_{\rm wind}$ is the density
of the Be star wind and $v_{\rm rel}$ is the relative speed between the
accretor and wind)  develops  two peaks: the
highest peak corresponds to the periastron passage (highest density), while
the second peak occurs when the drop in the relative velocity $v_{\rm rel}$
compensates the decrease in density (because of the inverse cube dependence).
Mart\'{\i} \& Paredes (1995) have shown that during both peaks the accretion
rate is above the
Eddington limit and therefore one expects that matter is ejected twice
within the 26.496 days interval. 
Mart\'{\i} \& Paredes  have found 
that variations of Be star wind velocity produce a variation
in the orbital phase of the second peak. 
 At this second accretion peak
the compact object is  far enough away from the Be star, so that the inverse
Compton losses are small and electrons can propagate out of the orbital plane.
Then an expanding double radio source should be observed, which in fact has
been observed by VLBI and MERLIN. 
   \begin{figure}
   \centering
   \includegraphics[angle=-90,width=9.5cm]{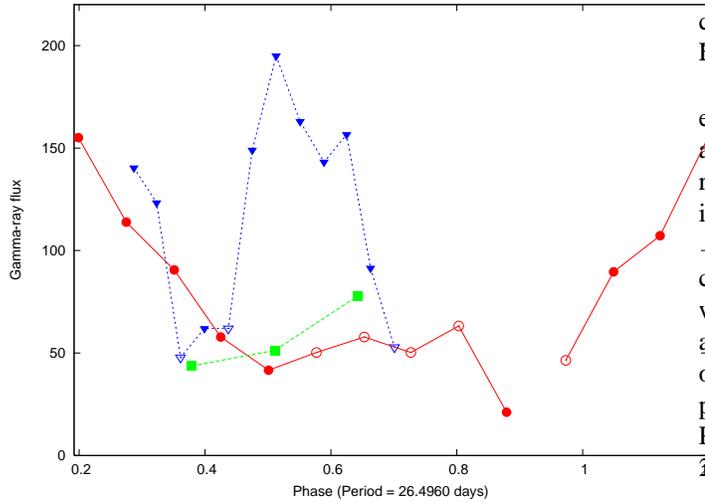}
      \caption{
% [ Massi et al. 2004b]
EGRET  data (Tavani et al. 1998) folded with the orbital/radio period
$P=26.4960$ days. 
 The plot begins at 
periastron passage $\Phi\simeq$0.2 and shows the follow-up of the gamma-ray
emission  along one full orbit. At  epoch 2450334JD 
(i.e. circles in the plot, with empty circles indicating  upper limits) 
the orbit has been well sampled at all phases: A clear peak is
centered at periastron passage 0.2 and 1.2.
At a previous epoch (2449045JD; triangles in the plot, with empty triangles
 indicating  upper limits) the sampling is uncomplete,  still the data
show an increase  toward  ($\Phi \simeq $0.3) periastron passage and a 
peak at $\Phi\simeq$0.5. The 3 squares refer to a third epoch (2449471JD). 
The emission is suggested to be   produced via
inverse Compton scattering of stellar photons (Taylor et~al. 1992, 1996)
and of synchrotron photons (Bosch-Ramon \& Paredes 2004)
by the relativistic electrons of
the jet. The ejection is  predicted to occur twice along the orbit,
one always at periastron passage and the second at a varing orbital phase
(Mart\'{\i} and Paredes 1995).
Relativistic electrons ejected at $\Phi$=0.2 suffer severe Compton losses:
radio outbursts  never occur at periastron.  
Radio outbursts occur in the orbital phase interval
 0.45-0.95 (Paredes et al. 1990).
The gamma-ray peak at $\Phi\simeq0.5$  could be associated to 
a such  second ejection.
However, no radio data are available to calculate energy budget/losses of the
relativistic electrons
(figure in  Massi et al. 2004b).
              }
         \label{FigVibStab}
   \end{figure}

\section{A precessing  jet}

The first VLBI observation resolving the source, made by
Massi and collaborators (1993; $\Phi$=0.74)
\begin{figure}
\centering
\includegraphics[width=8.5cm]{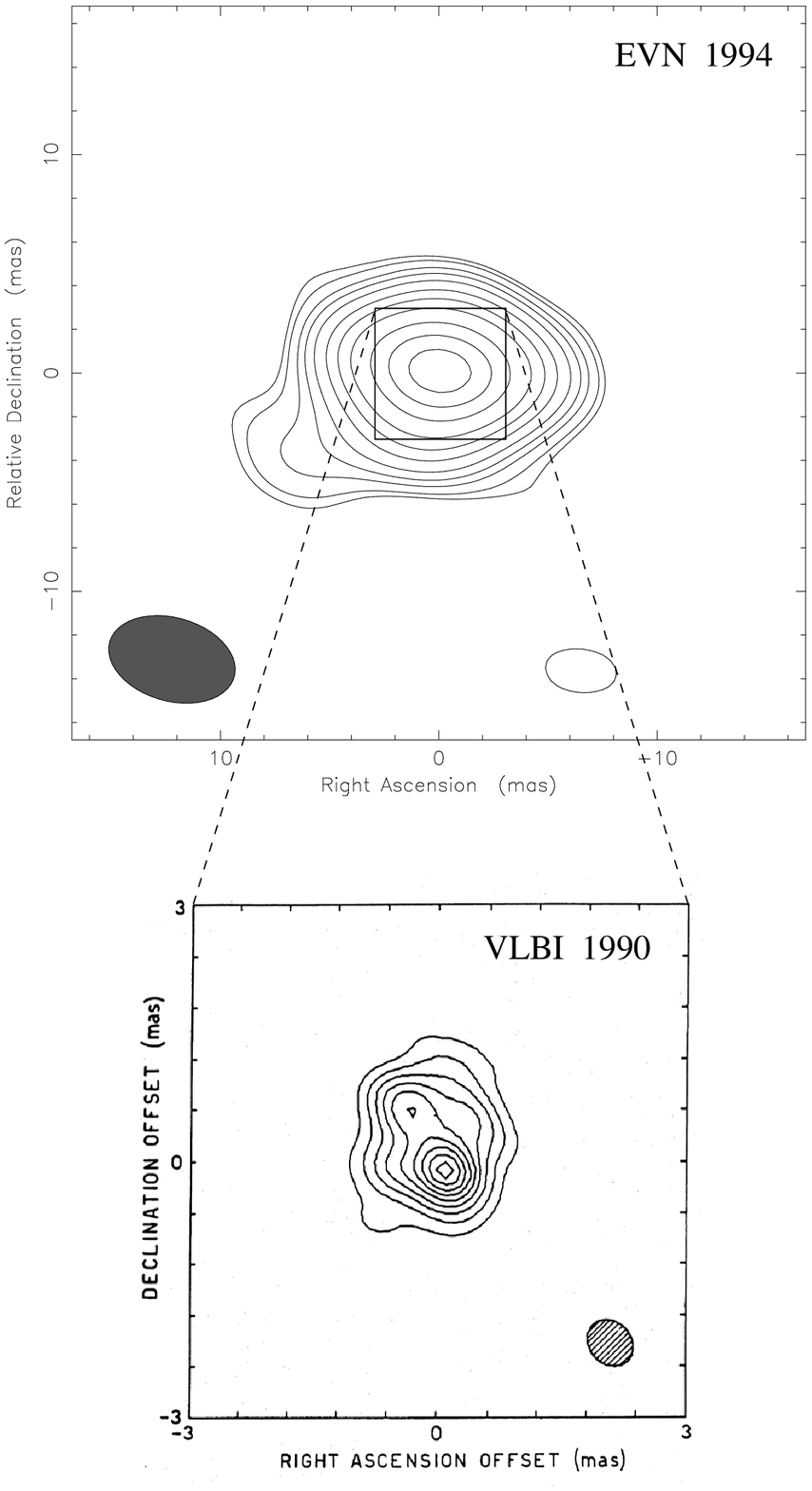}
\caption{
 Bottom: VLBI observation of LS~I~+61$^{\circ}$303 at 6cm.
 The telescopes involved were those at  Effelsberg  (Germany), Westerbork (Neth
erlands),
Medicina (Italy), Onsala (Sweden) and VLA (New Mexico, USA).
The observation, lasting  14 hours, was performed during a slow decay
of a large ($>250$mJy) outburst. The beam is 0.6 x 0.5 mas. The peak flux density is 36 mJy/beam. The lowest contour is 15$\%$ of the peak and the increment is of 10$\%$. Top:
EVN uniform weighted map of LS~I~+61$^{\circ}$303 at 6 cm.
The telescopes involved were those at  Effelsberg , Medicina , Noto (Italy) and Onsala.
  The contours are at $-$3, 3, 4, 6, 8, 11, 16, 22, 30, 45, 60 and 75 times the r.m.s. noise of 0.28~mJy~beam$^{-1}$. The filled ellipse in the bottom-left corner represents the FWHM of
the synthesized beam, which is 5.9~mas$\times$3.8~mas at a P.A.
 of 74.2$^{\circ}$ 
(figure in Massi et al. 2001).}
\label{fig:vlbi+evn}
\end{figure}
revealed a complex morphology (Fig.\ref{fig:vlbi+evn}~bottom)
 not easy to interprete.
Two components (at
P.A.$\simeq$30$^{\circ}$) are separated by 0.9 mas, 
corresponding to 1.8 AU at a distance of 2 kpc.
The two components are inside an  extended and sensibly rotated
 structure at  P.A.$\simeq$ 135$^{\circ}$).
 What is the nature of this  complex morphology ?
Is the  envelope an older expanding
jet, previously ejected, because of precession, at another angle?
Taylor and collaborators (2000; $\Phi$=0.69)
 performing VLBI observations
in combination with the HALCA orbiting antenna  
mapped a curved structure of 4 mas (8 AU) reminiscent
 ''of the precessing radio jet seen in SS433''.

EVN observations (Massi et al 2001; $\Phi$=0.92) 
at a scale of  up to tens of AU
show an elongation clearly in one
direction  without any ambiguity (see Fig. 4-Top).
The observed flux
density of the approaching ($S_{\rm a}$) and the receding ($S_{\rm r}$) jet
are a function of  $\theta$, the angle between the jet and the line of
sight,
by the Doppler factor: $\delta_{\rm a,r}=[ \Gamma (1 \mp
\beta\cos\theta)]^{-1}$, where $\Gamma=(1-\beta^2)^{-1/2}$ is the Lorentz
factor and $\beta~c$ the jet velocity (Mirabel \& Rodriguez 1999).
 The observed flux density of the
approaching jet  will be boosted and that of the
receding jets  de-boosted as
$S_{\rm a,r}=S \delta_{\rm a,r}^{k-\alpha}$, where $\alpha$ is the spectral
index of the emission ($S_{\nu}\propto \nu^{+\alpha}$) and $k$ is 2 for a
continuous jet and 3 for discrete condensations.
That creates an asymmetry 
between the two jet components. The receding attenuated jet can
even dissapear because of the sensitivity limit of the image. In this
case the jet appears only on one side as it is the case for the EVN image.

The first of two consecutive MERLIN observations  
(Massi et al. 2004; $\Phi$=0.68) shows a double  S-shaped jet
extending  to about 200~AU on both sides of a central source (Fig. 5 a).
The receding jet is attenuated but still above the noise limit of the  image. 

The precession suggested from the first MERLIN image becomes evident in the
second one ($\Phi$=0.71), shown in Fig.5~b, where a new feature is present
oriented to the North-East at a position angle (PA) of 67$\degr$. 
It is likely that the  morphology of the source is
S-shaped because the only visible  jet appears indeed bent.
The Northwest-Southeast jet of Fig.5~a has a PA=124$\degr$.
Therefore a quite large rotation has occurred in only 24 hours.

The  appearance of
successive ejections of a precessing jet with  ballistic
motion of each ejection 
is a curved path, that depending on the modality of the  expansion
and therefore on the adiabatic losses
seems  to be a  ``twin-corkscrew''  or a simply S-shaped pattern 
(Hjellming \& Johnston 1988; Crocker et al. 2002)
Can we distinguish in our data the single ejections in ballistic motion? 
 
We have split the MERLIN data of each epoch into blocks of a few hours  
and created separate images (Fig.~6) (i.e. Fig.~5a is the combination of the
first two blocks: 6a and 6b; Fig. 5b a combination of 6c and 6d).
 We see that the Eastern bent structure present in
Fig.~5a is the result of a combination of an old ejection A (Fig.~6a), already
displaced 120~mas from the core, and a new ejection B (Fig.~6b). After 19
hours (Fig.~6c) the feature B is reduced to $2\sigma$ and a new ejection C,
at a different PA with respect to B, appears. In Fig.~6d, 6 hours
later, little rotation of the PA is compatible with $\Delta$ PA$_{(\rm B-\rm
C)}/ 3$ of the previous image.

From the maps it is evident that  the projection of the jet
on the sky plane is changing, but how much is the variation
of $\theta$?
 If the jet velocity is the same for all ejections, the change of the ratio
${S_{\rm a}\over{S_{\rm r}}}=\left({1+\beta\cos\theta\over1-\beta\cos\theta}
\right)^{k-\alpha}$ is due to  variations  of  
$\theta$.
 Adopting
 values of $k=2$, $\alpha$=-0.5 and  the a value of $\beta=0.6$
we obtain:
$\theta_{\rm A} < 90\degr$, $\theta_{\rm B}< 80\degr$ and for
the C ejection in Fig.~6 c, $\theta_{\rm C} < 68 \degr$.

   \begin{figure*}
   \centering
   \includegraphics[width=12cm]{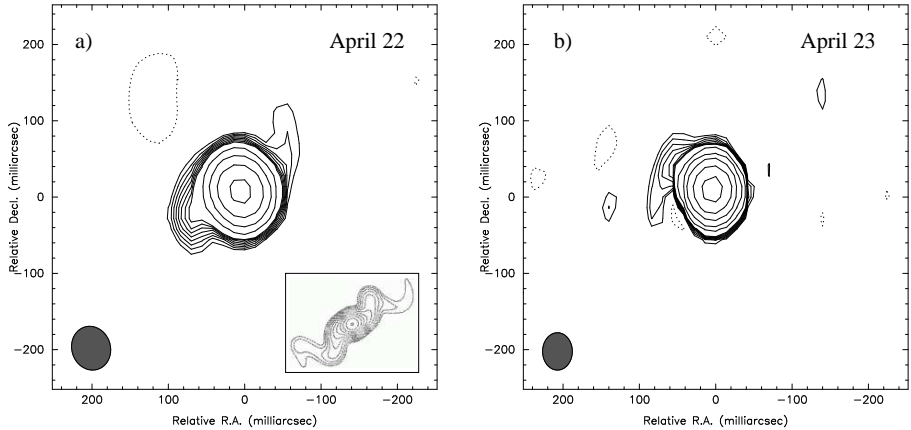}
      \caption{
{\bf a)} MERLIN  image of \object{LS~I~+61$^{\circ}$303} at
5~GHz  obtained on 2001 April 22. North is up and
East is to the left. The synthesized beam has a size of $51\times58$~mas, with
a PA of 17\degr. The contour levels are at $-$3, 3, 4, 5, 6, 7, 8, 9, 10, 20,
40, 80, and 160$\sigma$, being $\sigma$=0.14~mJy~beam$^{-1}$. The S-shaped
morphology strongly recalls the precessing jet of \object{SS~433}, whose
simulated radio emission (Fig.~6b in Hjellming \& Johnston \cite{hjellming88}
) is given in the small box. {\bf b)} Same
as before but for the April 23 run. The
synthesized beam has a size of $39\times49$~mas, with a PA of $-$10\degr. The
contour levels are the same as those used in the April 22 image but up to
320$\sigma$, with $\sigma$=0.12~mJy~beam$^{-1}$ (figure in Massi et al. 2004).
}
         \label{FigVibStab}
   \end{figure*}

   \begin{figure*}
   \centering
  \includegraphics[width=\textwidth]{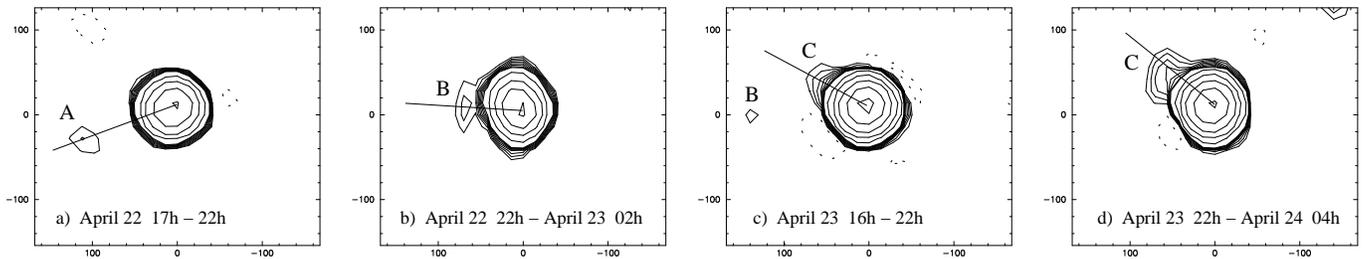}
\caption{
 MERLIN images of LS~I~+61$^{\circ}$303 of 2001 April 22 
and April 23. The data set of each epoch has been split
 into two blocks
(i.e. Fig.~5a is the combination of the
first two blocks: 6a and 6b; Fig. 5b a combination of 6c and 6d).
  A convolving beam of 40~mas has been used 
in all images for better display. The first contour represents 
the $3\sigma$ level in all images except for c), where we
 start from the $2\sigma$ level to display the faint B component. 
The rms noises are $\sigma$=0.13~mJy~beam$^{-1}$,
 $\sigma$=0.20~mJy~beam$^{-1}$, $\sigma$=0.13~mJy~beam$^{-1}$,
 and $\sigma$=0.15~mJy~beam$^{-1}$, respectively.
 The PA of the ejections is indicated by a bar (figure in Massi et al. 2004).}
         \label{FigVibStab}
   \end{figure*}

\section{Conclusions}

The main points of 
our review  on the characteristics 
of the  source LS~I~+61$^{\circ}$303 in the frame of the two-peak
accretion/ejection model 
are:

\begin{enumerate}
\item
It is still an open issue whether the
compact object in this system is a neutron star or a black hole.
In fact, taking into account the uncertainty in inclination,
mass of the companion and the mass function, the existence of a black
hole cannot be ruled out.

\item
The  radio jet  at a scale of hundreds of AU   quite strongly changes
its morphology in  short intervals (within 24 hours),  evolving
from an initial double-sided jet into  an one-sided
jet. This  variation
corresponds to a reduction of more than 10$^{\circ}$ in the angle between the jet and
the line of sight. This
new alignment severely Doppler de-boosts the counter-jet.
Further observational evidence for a precessing jet
is recognizable even  at AU scales.

\item
The same population
of relativistic electrons emitting
 radio-synchrotron radiation
may  upscatter - by inverse Compton processes -
 ultraviolet stellar photons  and produce gamma-ray emission.
For ejections at the periastron passage
gamma-ray flares are expected, but because of severe Compton losses
no radio flares, as indeed
the data seem to indicate.

\end{enumerate}

We conclude that  as precession
and variable doppler boosting
are   the causes of  the  rapid change in the radio-morphology,
 precession and variable  doppler boosting  are  likely
to produce    gamma-ray variations at short time scales.
The amplification due to the Doppler factor
for Compton scattering of stellar photons by the relativistic electrons of the
jet is $\delta^{3-2\alpha}$ (where $\alpha<0$), and therefore  higher than
that for synchrotron emission, i.e. $\delta^{2-\alpha}$ (Kaufman Bernad\'o
et~al. 2002).
 LS~I~+61$^{\circ}$303 becomes therefore the ideal laboratory to test
the recently proposed model for microblazars with INTEGRAL and MERLIN
observations now and by AGILE and GLAST in the future.

\begin{acknowledgements}

It is a pleasure to thank Karl Menten 
and J\"urgen Neidh\"ofer
 for careful reading of the
manuscript and  valuable  comments  and discussions.

\end{acknowledgements}

\end{document}